\documentclass[ aps,floatfix,  twocolumn, reprint, showpacs, nofootinbib]{revtex4-2}
\usepackage{graphicx}
\usepackage{amssymb}   
\usepackage{amsfonts}
\usepackage{amsmath}
\usepackage{dsfont}
\usepackage{dcolumn}   
\usepackage{bm}        
\usepackage[mathscr]{eucal}
\usepackage[dvipsnames]{xcolor}
\usepackage[colorlinks, linkcolor=blue,citecolor=blue,urlcolor=blue]{hyperref}
\usepackage[all]{hypcap} 
\usepackage{wasysym}

\begin{document}
	\title{Vacancy-induced local moments in  quantum paramagnetic phases: An SU($N$) designer Hamiltonian study}
	
	\author{Md Zahid Ansari}
	\thanks{These authors contributed equally to this work.}
	\affiliation{Department of Electrophysics, National Yang Ming Chiao Tung University, Science Building III, No 1001, University Rd., Hsinchu 300, Taiwan} 
	\author{Souvik Kundu}
	\thanks{These authors contributed equally to this work.}
	\affiliation{International Centre for Theoretical Sciences, Tata Institute of Fundamental Research, Bengaluru 560089, India}
	\author{Kedar Damle}
	\affiliation{Department of Theoretical Physics, Tata Institute of Fundamental Research, Mumbai 400 005, India} 
	%
	%
	\begin{abstract}
		
		We explore the effects of non-magnetic impurities (vacancy disorder) on the quantum paramagnetic phases stabilized by SU($N$) designer Hamiltonians on bipartite lattices. Using the results of our quantum Monte Carlo simulations, we demonstrate that isolated vacancies seed emergent spin $S=1/2$ moments in their vicinity when the low-temperature state has valence bond solid order. Indeed, our quantum Monte Carlo results for the low-temperature susceptibility in such regimes shows clear evidence of the vacancy-induced Curie tails associated with these emergent moments, and our zero-temperature projector Monte Carlo results on the ground-state wavefunction in the valence bond basis provide additional evidence in support of this picture. Further, for such designer Hamiltonians on the Lieb lattice with two additional sites on each bond of a square lattice, we identify a low-temperature spin liquid-like regime with no sign of spin or valence bond order. This liquid-like regime serves as a test bed for validating a recently-developed argument concerning the effects of vacancy disorder in such low temperature regimes. Consistent with this argument, we find that isolated vacancies do not seed emergent local moments in such spin liquids. Instead, in the presence of vacancy disorder, emergent local moments are associated with the presence of monomers in maximum-density dimer packings of the corresponding diluted lattice.
		
	\end{abstract}
	\maketitle
	
	
	\section{Introduction}

Spin systems, {\em i.e.}, insulators with interacting local moments, typically order at a temperature set by the scale of the magnetic interactions. Quantum spin liquids are unusual states that defy this trend, remaining in a liquid-like state down to the lowest temperature. Anderson's original proposal~\cite{Anderson_Materials_Research_Bulletin_1973,Fazekas_Anderson_The_Philosophical_Magazine_1974} for such a state was in the context of the triangular lattice Heisenberg antiferromagnet and couched in the language of valence bonds (singlets) between pairs of spins. In this proposal for a spin liquid, each spin forms a valence bond with a nearby spin, but the pattern of singlets has quantum fluctuations. The wavefunction of such a resonating valence bond (RVB) ground state is then a superposition of all possible ways for the valence bonds to form. 

Anderson's original variational construction of such a valence bond liquid state restricted the valence bonds to be between nearest-neighbor pairs of spins on the triangular lattice; however, having the valence bonds extend over a few lattice spacings (controlled by a length scale $\xi_{RVB}$) is not expected to qualitatively change the physics of such a short-range RVB (sRVB) spin liquid state~\cite{Liang_Doucot_Anderson_PRL_1988}. Such states of the spin system are expected to have a gap to both spin and singlet excitations.
Although this initial proposal of an RVB ground state on the triangular lattice was later shown to not be realized by the spin $S=1/2$ Heisenberg antiferromagnet~\cite{Huse_Elser_PRL_1988,Gelfand_Singh_Huse_PRB_1989,Singh_Huse_PRL_1992,Bernu_Lhuillier_Pierre_PRL_1992,Chubukov_Sachdev_Senthil_JPhysCondMat_1994,Capriotti_Trumper_Adolfo_Sorella_PRL_1999,White_Chernyshev_PRL_2007}, it nevertheless inspired a multitude of developments~\cite{Kivelson_Rokhsar_Sethna_PRB_1987,Rokhsar_Kivelson_PRL_1988,Moessner_Sondhi_PRL_2001,Kalmeyer_Laughlin_PRL_1987,Affleck_Marston_PRB_1988,Read_Sachdev_NuPhysB_1989,Read_Sachdev_PRL_1991,Sachdev_PRB_1992,Misguich_Lhuillier_Bernu_Waldtmann_PRB_1999,Wen_PRB_1991,Wen_QFTbook_2007,Wen_RevModPhys_2017,Moessner_Moore_TPMbook_2021,Sachdev_QPMbook_2023}, shaping the modern understanding of quantum many-body systems and the notion of topological order.

There has been considerable progress on the experimental front as well~\cite{Norman_RevModPhys_2016, Broholm_Cava_Kivelson_Nocera_Norman_Senthil_Science_2020, Clark_Abdeldaim_Annual_Review_of_Material_Research_2021}. One of the key difficulties in this regard is the fact that the most basic characterization of a spin liquid state is a negative one, in terms of the absence of any kind of ordering down to the lowest temperatures accessible to experiment. Apart from this lack of a definitive positive characterization in terms of an experimentally measurable signal, another complicating factor is the presence of quenched disorder, and the possible sensitivity of spin-liquid states to disorder effects~\cite{Broholm_Cava_Kivelson_Nocera_Norman_Senthil_Science_2020,
Clark_Abdeldaim_Annual_Review_of_Material_Research_2021,Lee_nature_2007,
Vries_Kamenev_Kockelmann_Benitez_Harrison_PRL_2008,Haussler_Sichelschmidt_2022,
Paddison_Daum_Dun_Ehlers_Liu_Stone_Zhou_Mourigal_Nature_Physics_2017}.

Gapped short-range RVB spin liquids are expected to be stable to weak bond randomness (exchange disorder), but strong exchange disorder is expected to destabilize the liquid ground state via a local moment instability~\cite{Kimchi_Nahum_Senthil_PRX_2018}. In contrast, valence bond solid states of quantum magnets, in which the singlet valence bonds display static order that breaks the symmetry of the underlying spatial lattice, are expected to be unstable to weak bond randomness~\cite{Kimchi_Nahum_Senthil_PRX_2018}. 

The response of gapped short-range RVB states to weak vacancy disorder, {\em i.e.}, dilution of the magnetic lattice by a small density of substitutional non-magnetic impurities, is expected to be quite different in the generic case: One expects weak vacancy disorder to lead to a local moment instability of such sRVB spin liquids except in certain special lattice geometries of which the kagome lattice is the example with the greatest potential experimental relevance~\cite{Ansari_Damle_PRL_2024,Bhola_Damle_arXivDec2025}. On the other hand, weak vacancy disorder is expected to generically lead to a local moment instability of valence bond solid states. 

This distinction between generic gapped sRVB liquid states and valence bond solid states has its origins in the following: In VBS states, each vacancy is expected to nucleate a local moment in its vicinity, as this allows the system to avoid the formation of an energetically-expensive domain wall in the valence-bond solid ordering. In contrast, vacancy-induced local moments in gapped sRVB states are a multi-vacancy effect associated with regions that host the monomers of any maximum-density dimer packing of the underlying diluted lattice~\cite{Ansari_Damle_PRL_2024,
Bhola_Biswas_Islam_Damle_PRX_2022,Bhola_Damle_arXivOct2023,Tseng_Anstee_2006,Aldred_Anstee_Locke_Discrete_Mathematics_2007, Anstee_Blackman_Yang_Discrete_Mathematics_2011}. In the generic case with a nonzero vacancy density, the number of such monomers in each connected component of the diluted lattice scales with its size, leading to a local moment instability. Exceptions include the kagome lattice, in which this number is at most $1$, thus precluding the possibility of any such local moment instability~\cite{Ansari_Damle_PRL_2024,Bhola_Damle_arXivDec2025}.

	These ideas have been tested~\cite{Ansari_Damle_PRL_2024} against numerical results from quantum Monte Carlo studies of designer Hamiltonians~\cite{Kaul_PRL_2015, Block_DEmidio_Kaul_PRB_2020} with enlarged symmetry, which realize a gapped quantum $\mathbb{Z}_{2}$ spin liquid phase on the kagome lattice, and a valence bond solid (VBS) phase on the triangular lattice.  On non-bipartite lattices such as the kagome and triangular lattices, this class of models has SO($N$) symmetry. For technical reasons having to do with the choice of sign-problem-free computational basis, it is not possible to obtain a direct quantum Monte Carlo estimate of the vacancy-induced susceptibility that corresponds to a field that couples to a generator of this symmetry. In Ref.~\cite{Ansari_Damle_PRL_2024}, this was finessed by using a surrogate observable that was diagonal in the computational basis and argued to also be sensitive to the presence of vacancy-induced Curie tails in the susceptibility.
	
If one could study designer Hamiltonians that realize VBS and sRVB phases of such models on bipartite lattices, one can go further: In the bipartite case, this class of designer Hamiltonians~\cite{Sandvik_PRL_2007,Lou_Sandvik_Kawashima_PRB_2009,Sandvik_PRL_2010,Melko_Kaul_PRL_2008,
Sandvik_2012_PRB,Pujari_Damle_Alet_PRL_2013,Kaul_Melko_Sandvik_2013,Kaul_PRB_2015,Pujari_Alet_Damle_PRB_2015,
Kaul_PRB_2012,Kaul_Sandvik_PRL_2012,Block_Melko_Kaul_PRL_2013,Harada_Kawashima_Troyer_PRL_2003,Jiang_2008,
Chen_2013,Kundu_Desai_Damle_2024} has SU($N$) symmetry. One of the additional symmetry generators is diagonal in the computational basis, allowing one to measure both the actual vacancy-induced susceptibility as well as the surrogate used earlier. One can therefore explicitly check that the surrogate observables indeed correctly mimic the low-temperature behavior of the susceptibility to a field that couples to this symmetry generator, as well as test the previous arguments about the contrasting nature of the vacancy-induced local moment instability of VBS and sRVB states.

	With this in mind, we consider SU($N$) designer models~\cite{Sandvik_PRL_2007,Lou_Sandvik_Kawashima_PRB_2009,Sandvik_PRL_2010,Melko_Kaul_PRL_2008,Sandvik_2012_PRB,Pujari_Damle_Alet_PRL_2013,Kaul_Melko_Sandvik_2013,Kaul_PRB_2015,Pujari_Alet_Damle_PRB_2015,Kaul_PRB_2012,Kaul_Sandvik_PRL_2012,Block_Melko_Kaul_PRL_2013,Harada_Kawashima_Troyer_PRL_2003,Jiang_2008,Chen_2013,Kundu_Desai_Damle_2024}, which are amenable to large-scale QMC simulations on bipartite lattices. In such models,  we can compute the two quantities: the ``ferromagnetic'' susceptibility and the ``antiferromagnetic'' susceptibility as defined in Sec. \ref{Methods and observables}. Of these, the ferromagnetic susceptibility, as the name suggests, is the SU($N$) analog of the actual susceptibility to a uniform field (that couples to the total spin $S^{z}_{\rm tot}$ in the SU($2$) case); in the SU($N$) case, it is the susceptibility to a field that couples to a SU($N$) generator that is diagonal in our computational basis. 
	
	The other ``antiferromagnetic'' suscepbitility is the response to a field that couples to the SU($N$) analog of the  antiferromagnetic Fourier component of $S^z(\vec{r})$ in the SU($2$) case; like in the SU($2$) case, this is not a symmetry generator. It is this second quantity that served as a surrogate for the physical (ferromagnetic) susceptibility in the earlier work~\cite{Ansari_Damle_PRL_2024} on O(N) designer Hamiltonians on the triangular and kagome lattice. The underlying argument was that this would also display an impurity-induced Curie tail whenever the physical (ferromagnetic) susceptibility had such an impurity-induced Curie tail at low temperature. Our work here thus allows for a direct verification of this underlying argument by measuring both kinds of responses in our simulations.
	
Our goal thus is to study both kinds of response in SU($N$) designer Hamiltonians that exhibit a VBS ordered ground state, and compare our findings to the behavior of such systems that instead exhibit a gapped SRVB ground state. There are several examples of VBS states of this type in the literature~\cite{Read_Sachdev_PRL_1989,Read_Sachdev_NuPhysB_1989,Read_Sachdev_PRL_1991,
Leung_Chiu_Runge_1996,Syljuaasen_2006,Ralko_Mambrini_Poilblanc_2009}. However, the same is not true for the sRVB states of interest to us. Indeed, there seems to be no well-established example of such a RVB ground state in such sign-free designer Hamiltonians that can be studied via large-scale quantum Monte Carlo simulations. Here, we circumvent this difficulty by identifying a finite-temperature, liquid-like regime in a tractable SU($N$) model defined on the $K=2$ Lieb lattice, {\em i.e.}, a decorated square lattice with $K=2$ sites added to lie on each bond of the original square lattice.

		The rationale for expecting such a liquid-like regime on the $K=2$ Lieb lattice is as follows:. In the limit $N \to \infty$, the designer Hamiltonian in question has an extensive ground-state degeneracy, which is expected to be lifted at finite $N$~\cite{Read_Sachdev_PRL_1989,Read_Sachdev_NuPhysB_1989}. These large-$N$ ground states are in correspondence with the fully-packed dimer covers of the underlying lattice, and perturbation theory in $1/N$ endows this dimer model with quantum dynamics, corresponding to a ring-exchange term on elementary plaquettes of the lattice~\cite{Read_Sachdev_PRL_1989,Read_Sachdev_NuPhysB_1989}.
		
		Such quantum dimer models are expected to have a crystalline ground state, which corresponds to a VBS state of the original magnet~\cite{Read_Sachdev_PRL_1989,Read_Sachdev_NuPhysB_1989}. However, the coefficient of the ring-exchange term that drives this ground-state ordering is expected to decrease exponentially with the number of bonds involved in this ring-exchange process, which in turn is set by the perimeter of the smallest plaquette of the lattice.  Now, this perimeter for the $K=2$ Lieb lattice is $12$ instead of the usual $4$ for a square lattice. We thus expect this ring-exchange term to be severely suppressed even for not-so-large values of $N$. Since the gap associated with the VBS ordering of the ground state is set by this microscopic energy scale, we expect the $K=2$ Lieb lattice to have a large liquid-like regime in temperature. It is this liquid-like regime that we exploit in this work.

	This paper is organized as follows. We discuss the SU($N$) designer Hamiltonians with two-site and multi-site interactions in Sec. \ref{Model}, followed by an overview of the method and the relevant quantities to be computed in QMC simulations in  Sec. \ref{Methods and observables}. We discuss the results of our numerical study in  Sec. \ref{Results}, with the concluding remarks presented in  Sec. \ref{Summary}.

	\section{Model}
	\label{Model}
We consider the following class of model Hamiltonians defined on a two-dimensional lattice:  
	\begin{align}
		H_{J} = -J_m\sum_{\langle ij \rangle} \hat{P}_{ij} \; , \label{SingletProjHam}
	\end{align}
	where the exchange coupling $J_m$ is taken positive and $\hat{P}_{ij}$ is defined as:	
	\begin{align}
		\hat{P}_{ij} = \frac{1}{N}\sum_{\alpha, \beta = 1}^{N} (\vert \alpha \rangle_i \vert \alpha \rangle_j) ( _i\langle \beta \vert  _j\langle \beta \vert )\; ,
	\end{align}
where the local Hilbert space at each site $i$ is $N$ dimensional, spanned by the orthonormal basis of ``color'' eigenstates $\vert \alpha\rangle_i$ (with allowed values $\alpha = 1 ,2 \dots N$).  

Clearly, this class of Hamiltonians has staggered global SU($N$) symmetry under the action of any SU($N$) rotation $U$ of the color states on all $A$ sublattice sites of the bipartite lattice, in conjunction with the action of its complex conjugate $U^{*}$ on all $B$ sublattice sites of the lattice. Thus, the color states on $A$ sublattice sites carry the fundamental representation of SU($N$) while those on the $B$ sublattice sites carry the complex conjugate of the fundamental representation~\cite{Read_Sachdev_NuPhysB_1989, Affleck_PRL_1985}.
From the point of view of this global symmetry, it is therefore clear that $P_{ij}$ projects on to the two-site singlet state $\frac{1}{\sqrt{N}}\sum_{\alpha} \vert  \alpha \rangle_i \vert \alpha \rangle_j $. 

Note that this interaction reduces (up to a constant) to the usual antiferromagnetic Heisenberg exchange interaction between two $S=1/2$ spins in the $N=2$ SU($2$) case.
On a non-bipartite lattice, one can again write down exactly the same class of model Hamiltonians. However, in this case, there is no bipartite sublattice structure, and the Hamiltonian has O($N$) symmetry, with color states on each site carrying the fundamental representation; in this non-bipartite case $P_{ij}$ is a projector that projects onto two-site O($N$) singlets~\cite{Kaul_PRL_2015, Block_DEmidio_Kaul_PRB_2020}.
	
	The minus sign in front of $J_m$ allows for a sign-free quantum Monte Carlo simulation of this class of designer Hamiltonians within the stochastic series expansion (SSE) approach. Indeed, such simulations are also possible with a variety of multi-projector terms included in the Hamiltonian. Without these additional multispin terms,  such systems with just the bilinear coupling $J_m$ typically exhibit a ground state with long-range color order (magnetic order) at small $N$, while the ground state at large values of $N$ is usually a quantum-disordered paramagnet that preserves SU($N$) or SO($N$) symmetry; in the bipartite case, this quantum-disordered ground state typically breaks lattice symmetries, corresponding to valence bond solid ordering~\cite{Harada_Kawashima_Troyer_PRL_2003, Kawashima_Naoki_Tanabe_PRL_2007, Beach_Alet_Mambrini_Capponi_PRB_2009, Lou_Sandvik_Kawashima_PRB_2009, Kaul_Sandvik_PRL_2012, Block_Melko_Kaul_PRL_2013, Kaul_PRL_2015, Block_DEmidio_Kaul_PRB_2020, Sandvik_PRL_2007, Melko_Kaul_PRL_2008, Sandvik_PRL_2010, Sen_Sandvik_PRB_2010, Banerjee_Damle_Paramekanti_PRB_2011, Pujari_Damle_Alet_PRL_2013, Pujari_Alet_Damle_PRB_2015, Iaizzi_Damle_Sandvik_PRB_2017, Iaizzi_Damle_Sandvik_PRB_2018}. 
	
	For instance, on the square lattice \cite{Harada_Kawashima_Troyer_PRL_2003}, the ground state has columnar VBS order for $N \geq 5$. As noted earlier, efficient simulations of the effects of multi-spin interactions are also possible as long as these are expressed as a product of projectors and satisfy some conditions on the signs of the corresponding coefficients.  Such multi-site interactions typically compete with the leading antiferromagnetic coupling $J_m$ and can drive the  transition to valence bond solid order to even at smaller values of $N$ \cite{Sandvik_PRL_2007, Kaul_PRL_2015}. With multi-column representations, {\em e.g.} for the SU($N$) analogs of $S=1$ antiferromagnets, such models can also realize a valence-bond nematic state on the square lattice for $N$ larger than a threshold value~\cite{Okubo_Harada_Lou_Kawashima_PRB2015,Kundu_Desai_Damle_2024}.

As the setting for our study of the effects of vacancies on VBS states, we focus on the the $S=1/2$ honeycomb-lattice $J\mbox{-}Q_3$ model    
	\begin{align}
		H &= H_{J} + H_{Q_{3}}, \; \; \text{where,}\\
		H_{Q3} &= -Q_{3}\sum_{\langle ijklmn \rangle} P_{ij}P_{kl}P_{mn} +  P_{jk}P_{lm}P_{ni}\; ,
		\label{eq:JQ3_model}
	\end{align} 
	where the sum in the second term is over hexagonal plaquettes $\langle ijklmn\rangle$ of the honeycomb lattice comprising these six sites listed in clockwise order, and the three projectors in a given term thus act on alternating bonds of that plaquette. 
$H_{Q_3}$ realizes a columnar VBS phase for $Q_{3} / J_m \gtrsim 1.19$~\cite{Pujari_Damle_Alet_PRL_2013, Pujari_Alet_Damle_PRB_2015}. 

In order to contrast these effects with the vacancy-induced response in a spin-liquid regime at low temperature, we study in parallel the singlet projector Hamiltonian $H_{J}$ on the $K=2$ Lieb lattice described earlier (in which each link of the square lattice hosts $K=2$ additional sites). We have checked (as we report in Sec.~\ref{Results}) that this designer Hamiltonian does indeed display such a liquid-like regime at moderately large values of $N$.
    
Our goal here is to study the vacancy-induced response in these low temperature regimes in two different situations. The first is that of isolated vacancies. We study this by removing two sites from an otherwise perfect lattice of a given finite size, with one of the sites being a $A$-sublattice site and the other belonging to the opposite $B$ sublattice. We ensure that the locations of the sites being removed are as far away from each other as possible.   

The second situation models the occurence of multi-vacancy clusters that trap the monomers of any maximum-density dimer packing~\cite{Lovasz_Plummer_book} of the diluted lattice. In a randomly-diluted disordered lattice, such clustering is rare. However, using the theoretical ideas developed earlier~\cite{Sanyal_Damle_Motrunich_PRL_2016,
Bhola_Biswas_Islam_Damle_PRX_2022,Bhola_Damle_arXivOct2023}, we can place some vacancies in specific positions relative to each other to create the simplest examples of such multi-vacancy clusters. 

The corresponding regions of the lattice (that trap the momomers of the dimer packing) have been dubbed ${\mathcal R}$-type regions in this earlier work, and we place two such ${\mathcal R}$-type regions in the sample, as far away from each other as possible. One of these regions is created by removing $A$-sublattice sites in a specific pattern and traps a single monomer of any maximum-density dimer packing of the disordered lattice on the $B$-sublattice sites within that region. The other region is chosen by reversing the roles of the $A$ and the $B$ sublattice in this construction.

To summarize: We investigate the impurity response in both scenarios by explicitly constructing the simplest vacancy patterns representative of each case. To model the first scenario, we introduce two isolated, maximally separated vacancies, one on each sublattice. To study the second scenario, we construct a pair of the simplest possible $\mathcal{R}$-type regions on the $K=2$ Lieb lattice, also placed at maximum separation on opposite sublattices, as illustrated in Fig.~\ref{fig:K2latticeRtype}.

	\section{Methods and observables}
	\label{Methods and observables}
	
	We use a recently developed color-resummed variant (RSSE)~\cite{Desai_Pujari_PRB_2021} of the stochastic series expansion (SSE)  algorithm~\cite{Sandvik_JPhysA_1992, Sandvik_PRB_1999, Syljuasen_Sandvik_PRE_2002, Sandvik_AIP_Conference_2010,Sandvik_Evertz_PRB_2010, Banerjee_Damle_JStatMech_2010} to perform quantum Monte Carlo (QMC) simulations of the finite temperature equilibrium properties of the SU($N$) designer Hamiltonians described in the previous section. Since we are focusing only on the paramagnetic regime of the SU($N$) systems where the average length of the operator loops in the deterministic loop update of SSE is very small, RSSE provides somewhat more efficient sampling, with improved Monte Carlo (MC) autocorrelation time compared to the standard SSE. For representative values of parameters, we have explicitly checked that results obtained using both algorithms match within statistical errors. We use approximately $\mathcal{O}(10^5)$ MC steps for warm up in the SU($2$) honeycomb $J\mbox{-}Q_3$ model. However, for large $N$, it takes longer to warm up. For example, $\sim\mathcal{O}(10^6)$ warm-up steps in the case of SU($8$) pure-$J$ model on the $K = 2$ Lieb lattice. 
	
	A complementary approach, which provides information about the ground state wavefunction in the bipartite valence bond basis~\cite{Lieb_1989,Beach_Sandvik_2006}, is the valence-bond projector (VBP) quantum Monte Carlo algorithm~\cite{Sandvik_Evertz_PRB_2010}. We have used this to obtain complementary information that provides a more detailed check on the theoretical arguments of Ref.~\cite{Ansari_Damle_PRL_2024}. To ensure convergence for the honeycomb VBS phase, we have used a Monte Carlo projection length of $6 \times L^{3}$. As alluded to in the introduction, the $K = 2$ Lieb lattice is expected to have a very small energy gap, necessitating a very high projection length in any projector method. Nevertheless, we have checked that our results for the $K = 2$ Lieb 	
	lattice using two different projection lengths ($6 \times L^{3}$, $8 \times L^{3}$) are within statistical errors.

		The most direct way to detect the presence of vacancy-induced local moments in our nonzero temperature simulations is of course to compute $ \chi^{\mathcal{Q}}$, the SU($N$) analog of the ferromagnetic susceptance ({\em i.e.}, the extensive analog of the susceptibility). This is the susceptance corresponding to a static external field that couples to $\mathcal{Q}_{\alpha\alpha}^{\rm tot}$  = $ \sum_{\vec{r}} (-1)^{r}\mathcal{Q}_{\alpha\alpha}(\vec{r}) $,  a diagonal generator of the staggered global SU($N$) symmetry of the system. Here, $\mathcal{Q}_{\alpha\alpha}(\vec{r})  = (|\alpha \rangle_r)( _r\langle \alpha \vert)
		 - \frac{1}{N}$ is the corresponding diagonal generator of the onsite SU($N$) symmetry and the staggering factor $(-1)^r$, which is $+1$ ($-1$)for all $A$ ($B$) sublattice sites, keeps track of the fact that spins on the $A$ sublattice sites carry the fundamental representation while those on the $B$ sublattice site carry the complex conjugate of the fundamental representation.
		
		In addition, we also compute $ \chi^{n}$, the SU($N$) analog of the antiferromagnetic susceptance.  This is the susceptance corresponding to a static external field that couples to $n^{\rm tot}_{\alpha\alpha}$  = $ \sum_{\vec{r}} \mathcal{Q}_{\alpha\alpha}(\vec{r}) $, an SU($N$) analog of the antiferromagnetic order parameter. For the designer Hamiltonian like Eq.~\ref{SingletProjHam} on a non-bipartite lattice, $n^{\rm tot}_{\alpha\alpha}$ represents an O($N$) analog of the nematic order parameter for the O($3$) case~\cite{Kaul_PRL_2015, Block_DEmidio_Kaul_PRB_2020, Ansari_Damle_PRL_2024}. 
	
In order to avoid any confusion regarding our normalization, it is perhaps useful to be a bit more explicit about the definition of these quantities: 	The static susceptance $\chi^{\hat{\mathcal{O}}}$ corresponding to an  operator  $\hat{\mathcal{O}}^{\rm tot}_{\alpha\alpha}$ (where $\hat{\mathcal{O}}^{\rm tot}_{\alpha\alpha}$ can be  $\mathcal{Q}_{\alpha\alpha}^{\rm tot}$ or $n^{\rm tot}_{\alpha\alpha}$) is:
	\begin{align}
		\chi^{\hat{\mathcal{O}}} = \frac{1}{N} \sum_{\alpha}\int_{0}^{\beta} \big\langle \hat{\mathcal{O}}_{\alpha\alpha}^{\rm tot}(  \tau)\hat{\mathcal{O}}_{\alpha\alpha}^{\rm tot}( 0) \big \rangle d\tau,
	\end{align}
	with $\hat{\mathcal{O}}_{\alpha\alpha}^{\rm tot}(\tau) = e^{\tau H} \hat{\mathcal{O}}_{\alpha\alpha}^{\rm tot} e^{-\tau H}$.  
	
We compute these quantities for a pure system of a certain size, and compare the results with those obtained after introducing the vacancies. The vacancy-induced impurity susceptibility is then the difference of the two results: 
		\begin{align}
		\chi_{\rm imp}^{\mathcal{Q}} &= \chi_{\rm disordered}^{\mathcal{Q}} - \chi_{\rm pure}^{\mathcal{Q}} \nonumber\\
		\chi_{\rm imp}^{n} &= \chi_{\rm disordered}^{n} - \chi_{\rm pure}^{n}.
		\label{eq:Impurity suceptibility}
	\end{align}
	
	Our  zero-temperature projector QMC simulations provide complementary information, in the sense that we have direct access to some of the properties of the ground state itself. In particularly, we are in a position to study which spins are paired with each other by valence bonds with a large amplitude in the ground state wavefunction. The idea here is the following: Assume emergent local moments form at intermediate energy-scales, with a spatial profile that is tightly-localized in the immediate vicinity of the two defects introduced into the system. Since the pure system is in a gapped state (either a VBS-ordered or sRVB spin liquid state), one then expects the structure of the ground state to remain largely undisturbed away from the two defects. In addition, one expects the two emergent local moments to interact antiferromagnetically with each other via a weak exchange coupling mediated by the bulk.
	As a result, one expects the two emergent local moments to be bound into a singlet state in the ground state wavefunction, although they are located far away from each other.
	
	In terms of the microscopic variables and their ground state wavefunction, this translates to the expectation that the ground state will feature a large amplitude for valence bonds that connect the immediate neighborhood of one defect with the immediate neighborhood of the other defect. More precisely, one expects a large amplitude for spins on the $B$ sublattice sites in the immediate vicinity of a $A$-sublattice defect to form singlets with the spins on $A$ sublattice sites in the immediate vicinity of the $B$-sublattice defect, although the two defects are placed far apart. This serves as a clear signature of a local-moment instability.

	\section{Results}
	\label{Results}
	
	\subsection{Finite temperature}\label{subsec:finite_temp}
	\begin{figure}[t]
    \begin{tabular}{cc}
        \includegraphics[width=0.5\columnwidth]{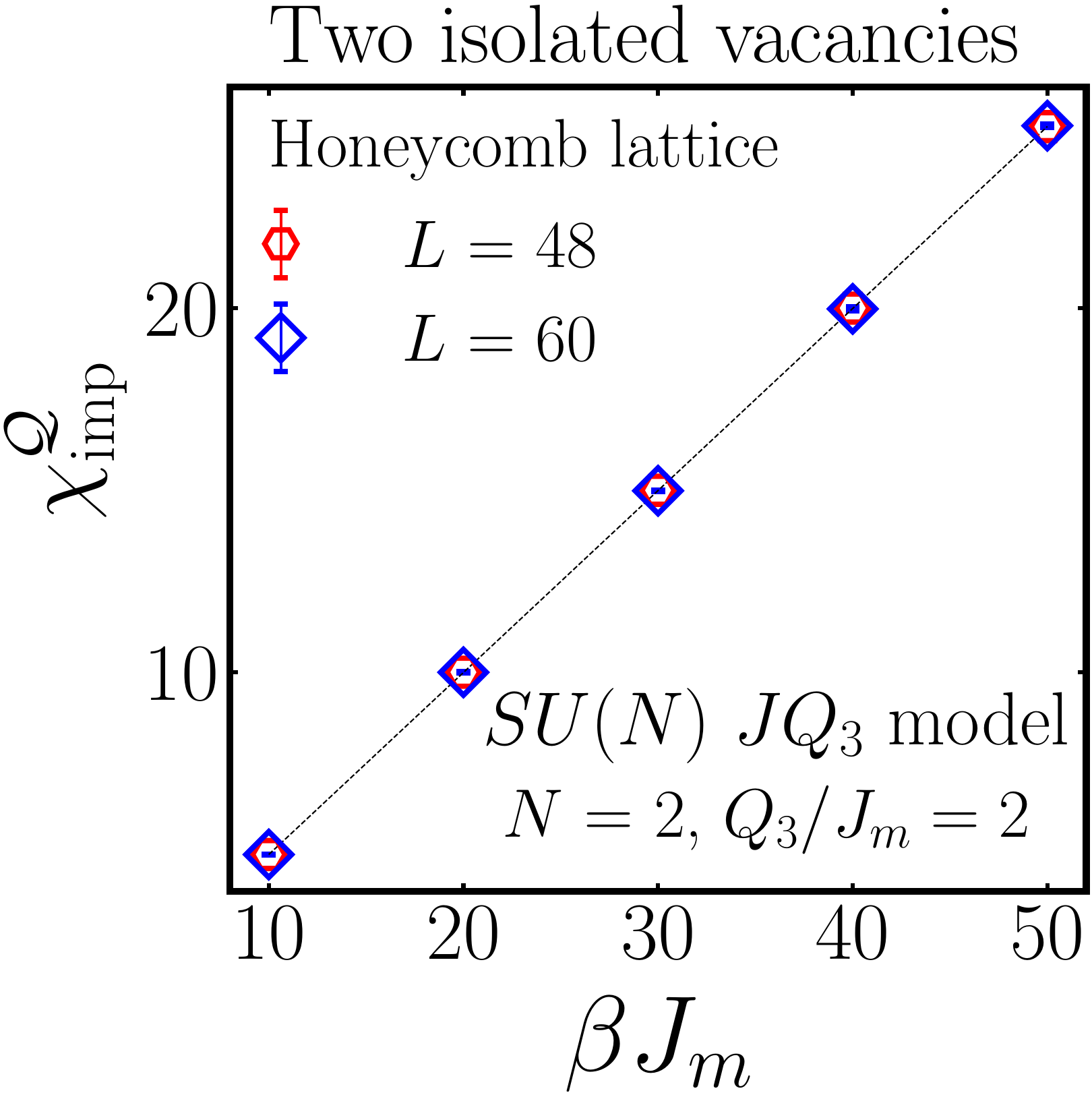} &
        \includegraphics[width=0.5\columnwidth]{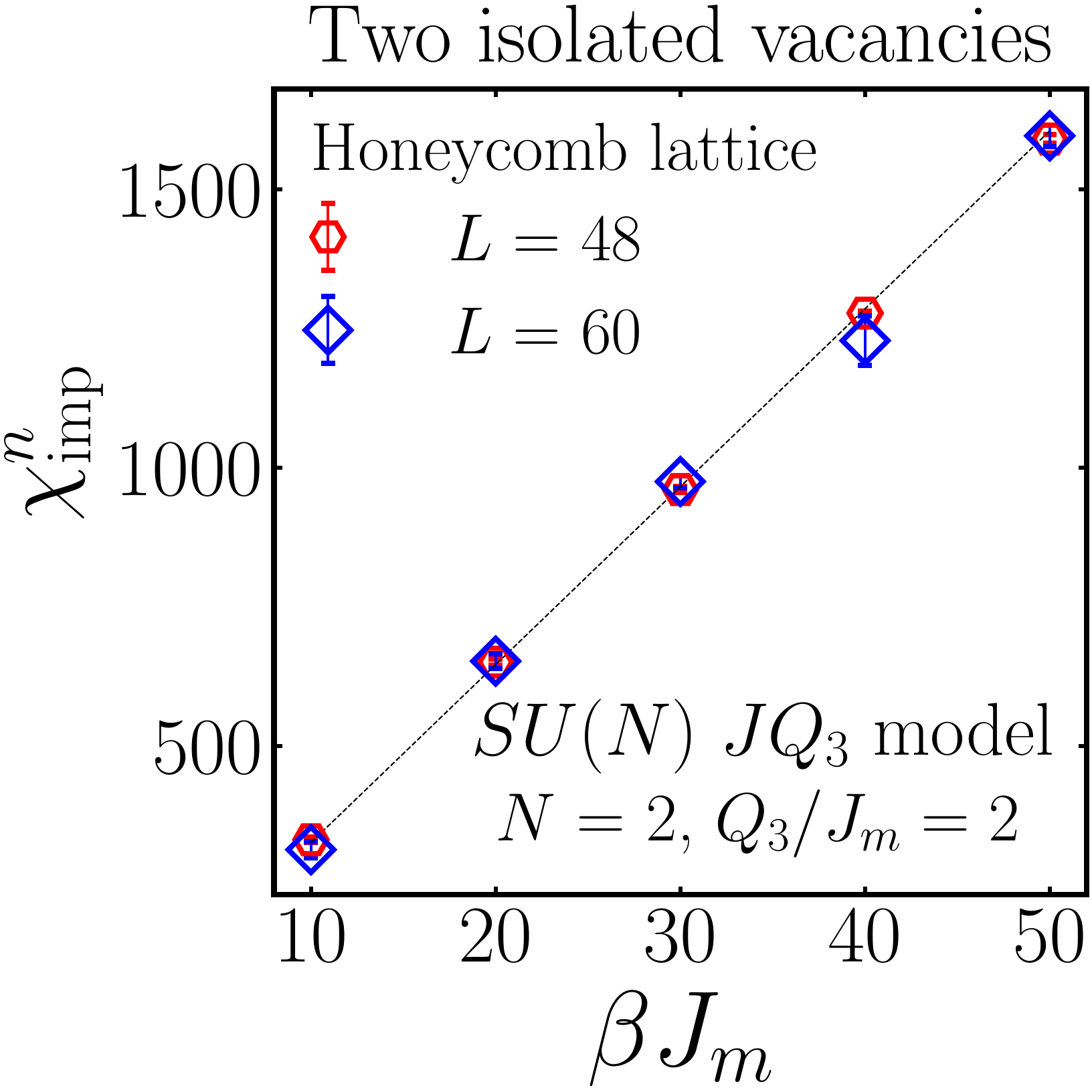}
    \end{tabular}		
    \caption{Impurity susceptibilities associated with the response to uniform static fields that couple to  $\mathcal{Q}_{\alpha\alpha}^{\rm tot}$ and $n^{\rm tot}_{\alpha\alpha}$ respectively, for two well-separated vacancies in the columnar VBS phase of the honeycomb-lattice $SU(2)$ $J\mbox{-}Q_3$ model. The data are obtained for $J_m=1.0$ and $Q_3=2.0$ on an $L \times L$ honeycomb lattice (with $L^2$ unit cells) with periodic boundary conditions, containing two isolated vacancies placed on opposite sublattices and separated by a distance $L/2$. Left panel: Data for $\chi_{\rm imp}^{\mathcal{Q}}$ reveals the presence of a clear low-temperature Curie tail, 
    	$\chi_{\rm imp}^{\mathcal{Q}} \propto 1/T \equiv \beta$, 
    	consistent with the presence of two effectively free 
    	SU($2$) moments induced by the vacancies. The solid line is a linear fit whose slope matches
    	the Curie constant of free spin-$\frac{1}{2}$ moments. Right panel: Impurity susceptibility 
    	$\chi_{\rm imp}^{n}$ also displays Curie-like behavior over the same temperature range, 
    	indicating that the vacancy-induced moments contribute to both uniform and staggered channels in the VBS phase. 
    }
    \label{fig:HoneycombdoulbleVac}
\end{figure}

\begin{figure}[t]
    \begin{tabular}{cc}
        \includegraphics[width=0.45\columnwidth]{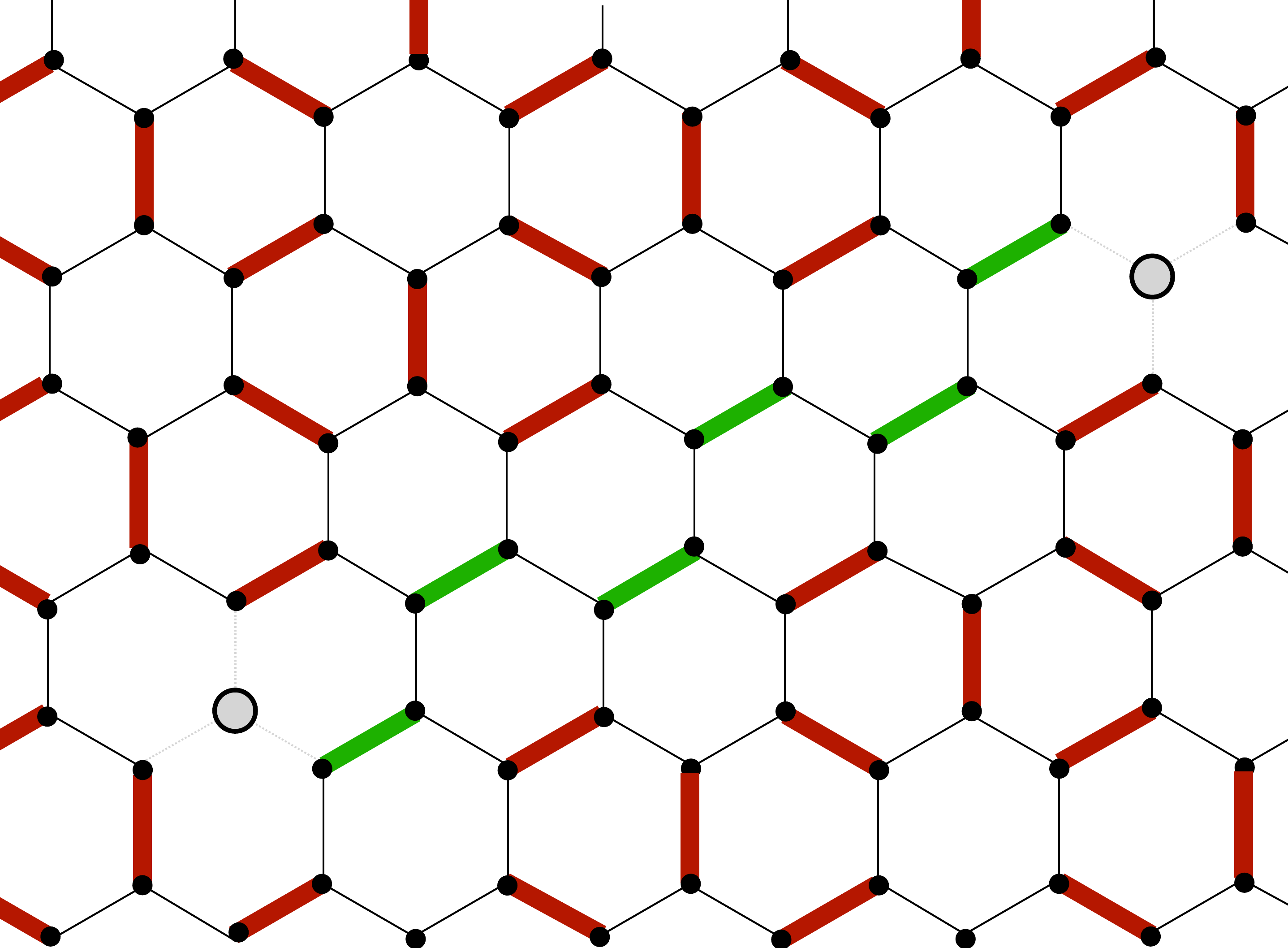} \hspace{0.5em} &
        \hspace{0.5em} \includegraphics[width= 0.45\columnwidth]{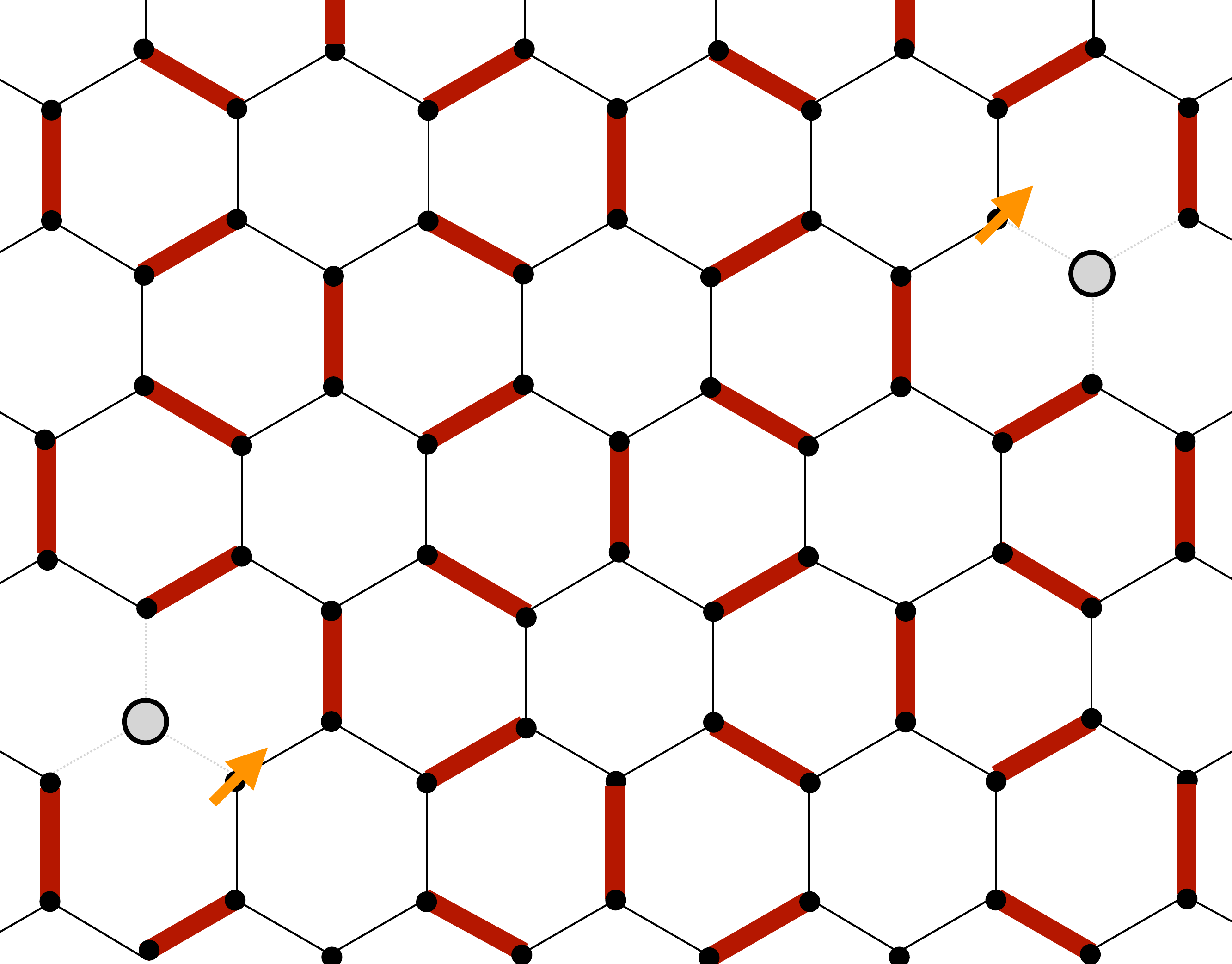}
    \end{tabular}
    \caption{Right panel: Schematic representation of a columnar valence bond solid ordered state (valence bonds are represented by thick red bonds) on the honeycomb lattice, with two isolated vacancies (represented by gray circles). Each vacancy nucleates a localized free spin-$1/2$ moment (indicated by yellow arrows) on neighboring sites. Left panel: In the presence of the two vacancies, a singlet state that tries to minimizes local distortions to the columnar VBS pattern unavoidably features a domain wall (color coded green). The domain wall separates two symmetry-related columnar VBS patterns on either side and incurs an energy cost proportional to its length in the columnar VBS phase.}
    \label{fig:honeycombDomainWall}
\end{figure}

We first probe the effect of two well-separated single-site vacancies in a valence bond solid (VBS) phase. To access a robust VBS state, we consider the SU($2$) $J\mbox{-}Q_{3}$ model on the honeycomb lattice [Eq.~\ref{eq:JQ3_model}] at $Q_{3}/J_m=2$, a parameter regime known to stabilize a three-sublattice symmetry-breaking columnar VBS phase~\cite{Pujari_Damle_Alet_PRL_2013}.

In Fig.~\ref{fig:HoneycombdoulbleVac}, we show the impurity susceptibilities $\chi^{\mathcal{Q}}_{\rm imp}$ (left panel) and $\chi^{n}_{\rm imp}$ (right panel) [defined in Eq.~\ref{eq:Impurity suceptibility}] as functions of $\beta J_m$ for two isolated vacancies placed on opposite sublattices and separated by a distance $L/2$. Both susceptibilities exhibit a clear Curie-tail behavior, $\chi_{\rm imp}\sim \beta$, at low temperatures ($\beta J_m\gg1$). In particular, the ferromagnetic impurity susceptibility $\chi^{\mathcal{Q}}_{\rm imp}$ quantitatively matches that of two free $S=1/2$ moments.

This behavior is consistent with energetic considerations in the VBS phase. The removal of a spin locally disrupts the singlet pattern, leading to the emergence of a localized moment near each vacancy in order to preserve the VBS order in the bulk. Although it is in principle possible to maintain a fully packed nearest-neighbor singlet covering without forming local moments, such configurations would require the introduction of a domain wall whose energy cost grows linearly with the separation between the two vacancies. This scenario is schematically illustrated in Fig.~\ref{fig:honeycombDomainWall}. We emphasize that the finite-temperature Curie response alone does not resolve the detailed spatial structure of these localized moments; this is evidenced indirectly through our zero-temperature analysis in Sec.~\ref{subsec:zero_temp}.

\begin{figure}[t]
    \centering
    \includegraphics[width=0.8\columnwidth]{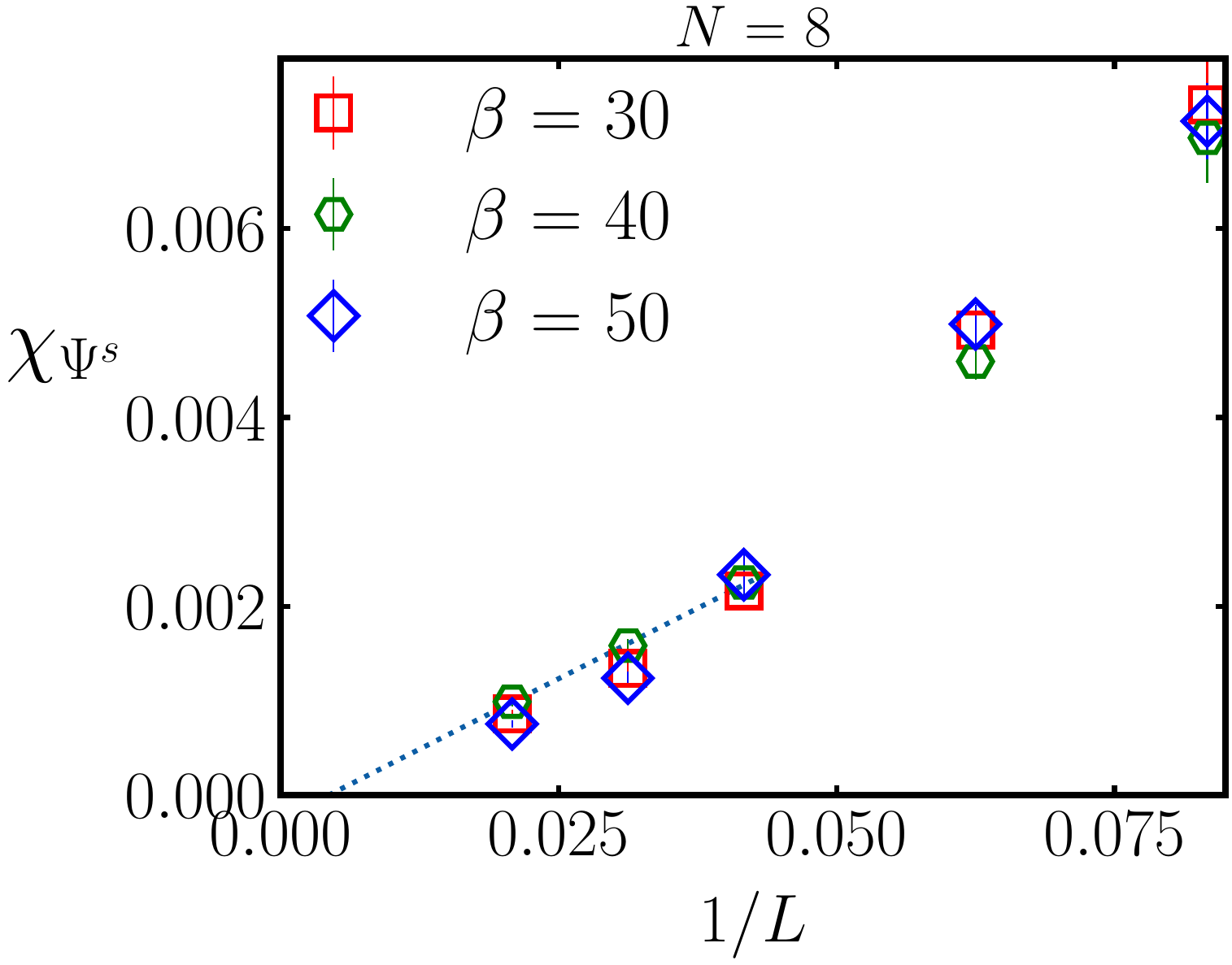}
    \caption{Susceptibility $\chi_{\Psi}$ associated with the VBS order parameter $\Psi$, defined in Eq.~\ref{eq:chi_vbs}, plotted as a function of $1/L$ for the $SU(N)$ nearest neighbor singlet projector Hamiltonian on the $K=2$ Lieb lattice at $N=8$ and several low-temperature (large $\beta$) values. The susceptibility decreases systematically with increasing system size and extrapolates to zero in the thermodynamic limit, indicating the absence of long-range columnar VBS order at all temperatures studied.  }
    \label{fig:structurefactorK2}
\end{figure}

Having established the impurity response in a VBS phase, we now turn to the corresponding question in an SU($N$) liquid. As discussed in the introduction, staggered SU($N$) interactions on bipartite lattices generically favor VBS order in the large-$N$ limit~\cite{Read_Sachdev_NuPhysB_1989}. This can be understood within an effective Hamiltonian framework: at $N=\infty$, the staggered SU($N$) model maps onto a classical dimer model, where each dimer represents a nearest-neighbor SU($N$) singlet~\cite{Read_Sachdev_NuPhysB_1989}. To leading order in $1/N$, quantum fluctuations lift the extensive degeneracy via an effective Hamiltonian acting on elementary flippable plaquettes. The resulting ground state maximizes the number of such plaquettes, leading to a columnar VBS phase. On the square lattice, this yields a fourfold-degenerate columnar VBS, in agreement with large-scale quantum Monte Carlo studies~\cite{Kawashima_Naoki_Tanabe_PRL_2007}.

By analogy, the staggered SU($N$) model on the $K=2$ Lieb lattice is expected to support a similar columnar VBS order at wave vectors $\mathbf{k}_{1}=(\pi,0)$ and $\mathbf{k}_{2}=(0,\pi)$ of the underlying square Bravais lattice. However, in this case the leading-order effective Hamiltonian acts on a larger 12-site plaquette, resulting in an unusually small energy gap separating the columnar VBS ground state from competing configurations. As a consequence, the large-$N$ phase on the $K=2$ Lieb lattice may retain liquid-like characteristics down to very low temperatures.

To examine this possibility, we compute the susceptibility associated with the complex VBS order parameter,
\begin{align}
\Psi = \psi_x + i\psi_y \; \; , \; \; \nonumber
\psi_{x(y)} = \frac{1}{N_{\rm site}}\sum_{\vec r} P_{r_{x(y)}} (-1)^{r_{x(y)}},
\end{align}
where $\vec r = r_x \hat x + r_y \hat y$ denotes the Bravais lattice vector. The corresponding susceptibility is
\begin{equation}
\label{eq:chi_vbs}
\chi_{\Psi} = \int_{0}^{\beta} \langle \Psi(\tau)\Psi^{*}(0) \rangle d\tau .
\end{equation}

In Fig.~\ref{fig:structurefactorK2}, we present $\chi_{\Psi}$ for $N=8$ as a function of $1/L$ for several inverse temperatures $\beta$ (the system undergoes a transition from Néel order to a quantum paramagnet near $N=7$). The data clearly show the absence of long-range columnar VBS order in the thermodynamic limit, even at the lowest temperatures accessible to our simulations. We therefore conclude that the large-$N$ phase of the $K=2$ Lieb lattice behaves as a quantum liquid at low temperatures.

\begin{figure}[t]
		\begin{tabular}{cc}
			\includegraphics[width=0.5\columnwidth]{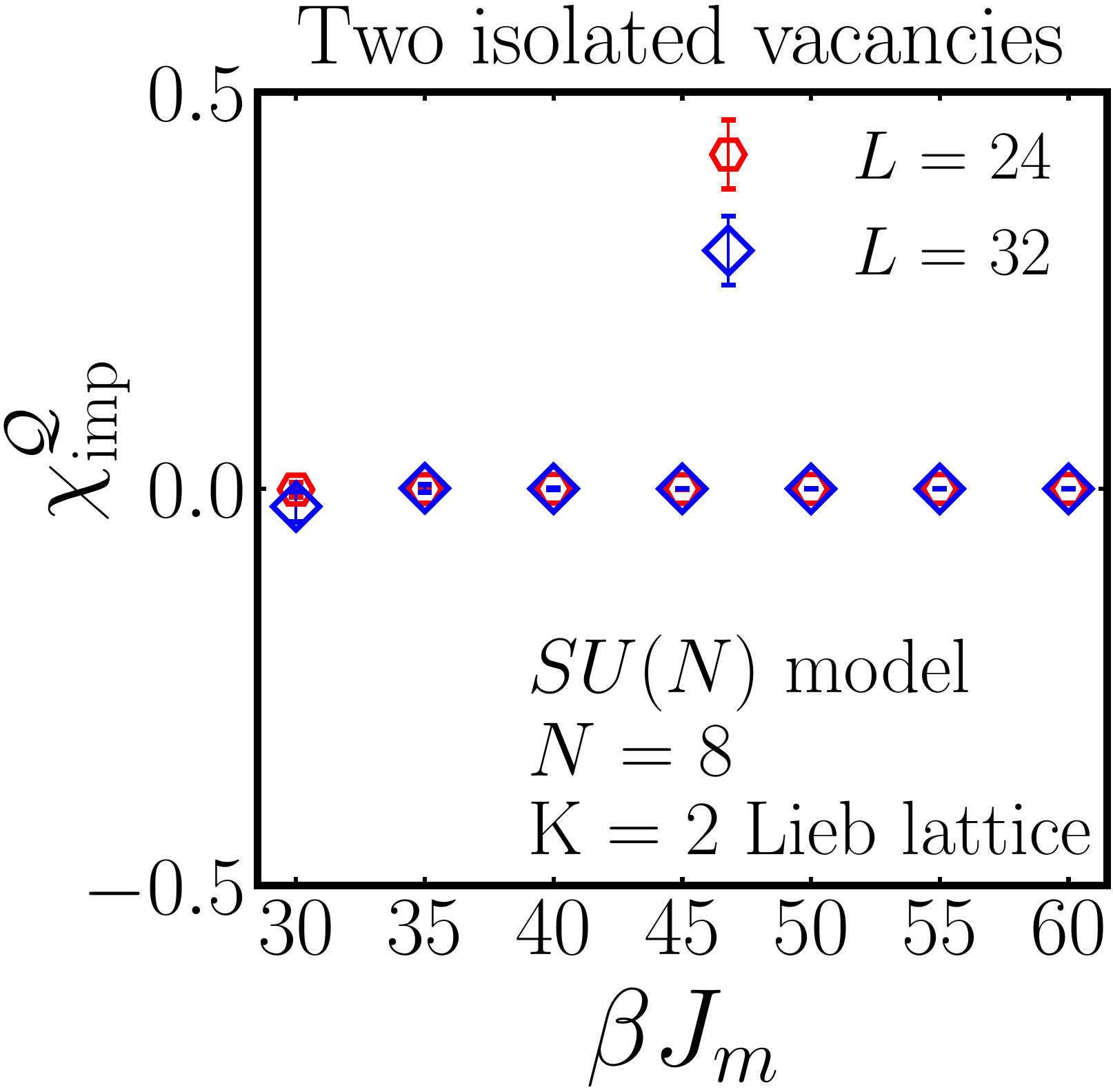} &
			\includegraphics[width= 0.5\columnwidth]{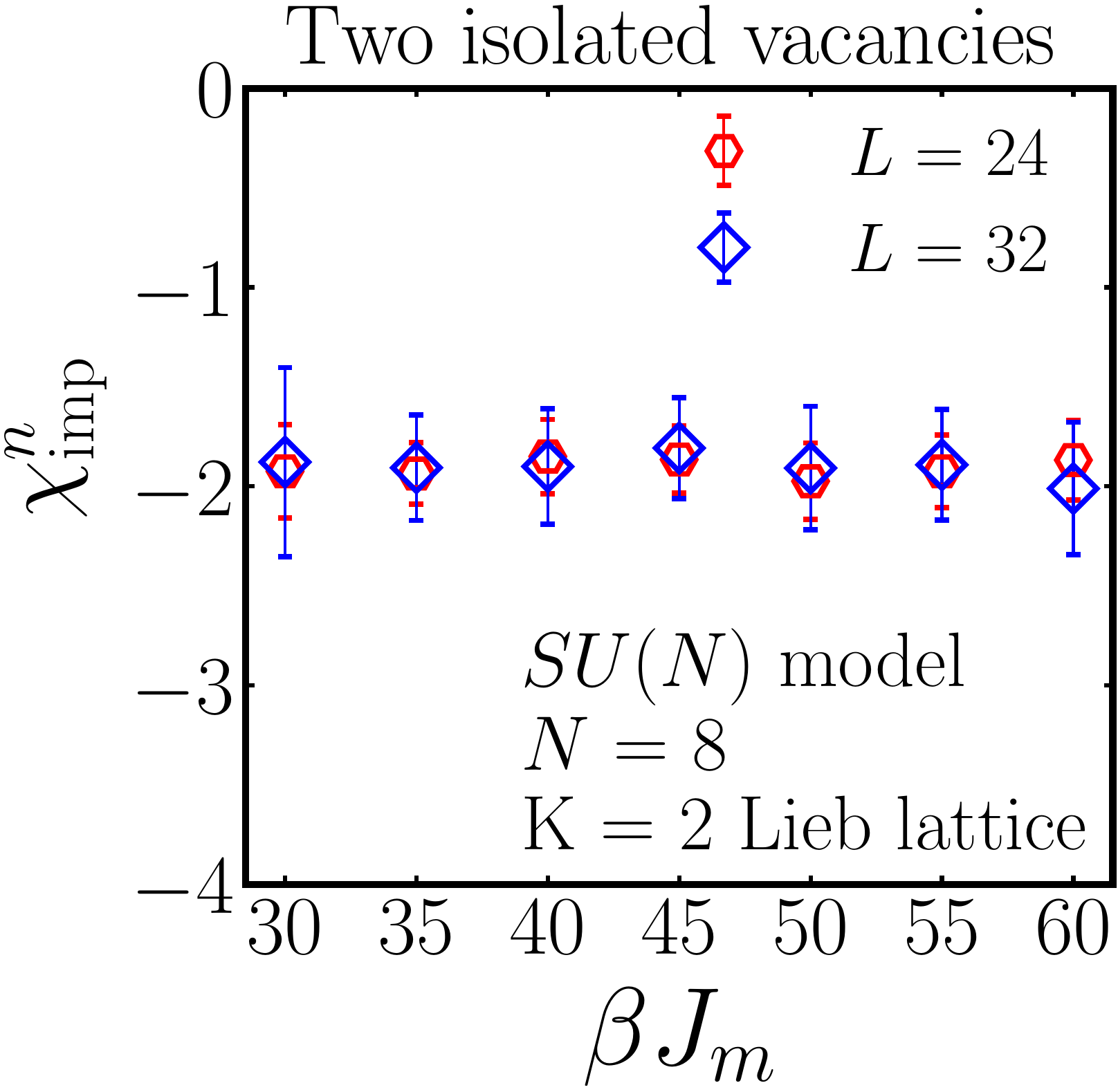}
		\end{tabular}		
		\caption[Impurity susceptibility corresponding to the uniform and alternating sum of the order parameter $\mathcal{Q}_{\alpha\alpha}$]{Impurity susceptibilities associated with the response to uniform static fields that couple to  $\mathcal{Q}_{\alpha\alpha}^{\rm tot}$ and $n^{\rm tot}_{\alpha\alpha}$ respectively. The data is displayed for two isolated vacancies separated by $L/2$ in an otherwise pure $L \times L$,  $K = 2$ Lieb lattice with $L^2$ unit cells and periodic boundary conditions imposed on both directions.
			Left panel: The impurity susceptibility $\rm \chi_{imp}^{\mathcal{Q}}$ (defined in the text) of the $K = 2$ Lieb lattice SU($N$) projector model at $N = 8$ shows absolutely no sign of a Curie tail at low temperature.
			Right panel: The impurity susceptibility $\rm \chi_{imp}^{n}$ (defined in the text) of the $K = 2$ Lieb lattice SU($N$) projector model at $N = 8$ also shows absolutely no sign of a Curie tail at low temperature.  
		}
		\label{fig:K2doubleVacancy}
	\end{figure}

We now investigate the impurity response in this liquid-like regime. We first consider two isolated vacancies placed on opposite sublattices and separated by a distance $L/2$. Figure~\ref{fig:K2doubleVacancy} shows the impurity susceptibilities corresponding to the uniform and staggered combinations of $\mathcal{Q}_{\alpha\alpha}$. Strikingly, neither susceptibility exhibits a Curie-like divergence, indicating the absence of localized ``spin'' degrees of freedom bound to isolated vacancies. This behavior stands in sharp contrast to the honeycomb VBS phase, where vacancies induce free local moments.

\begin{figure}[t]
		
		\includegraphics[width=1.0\columnwidth]{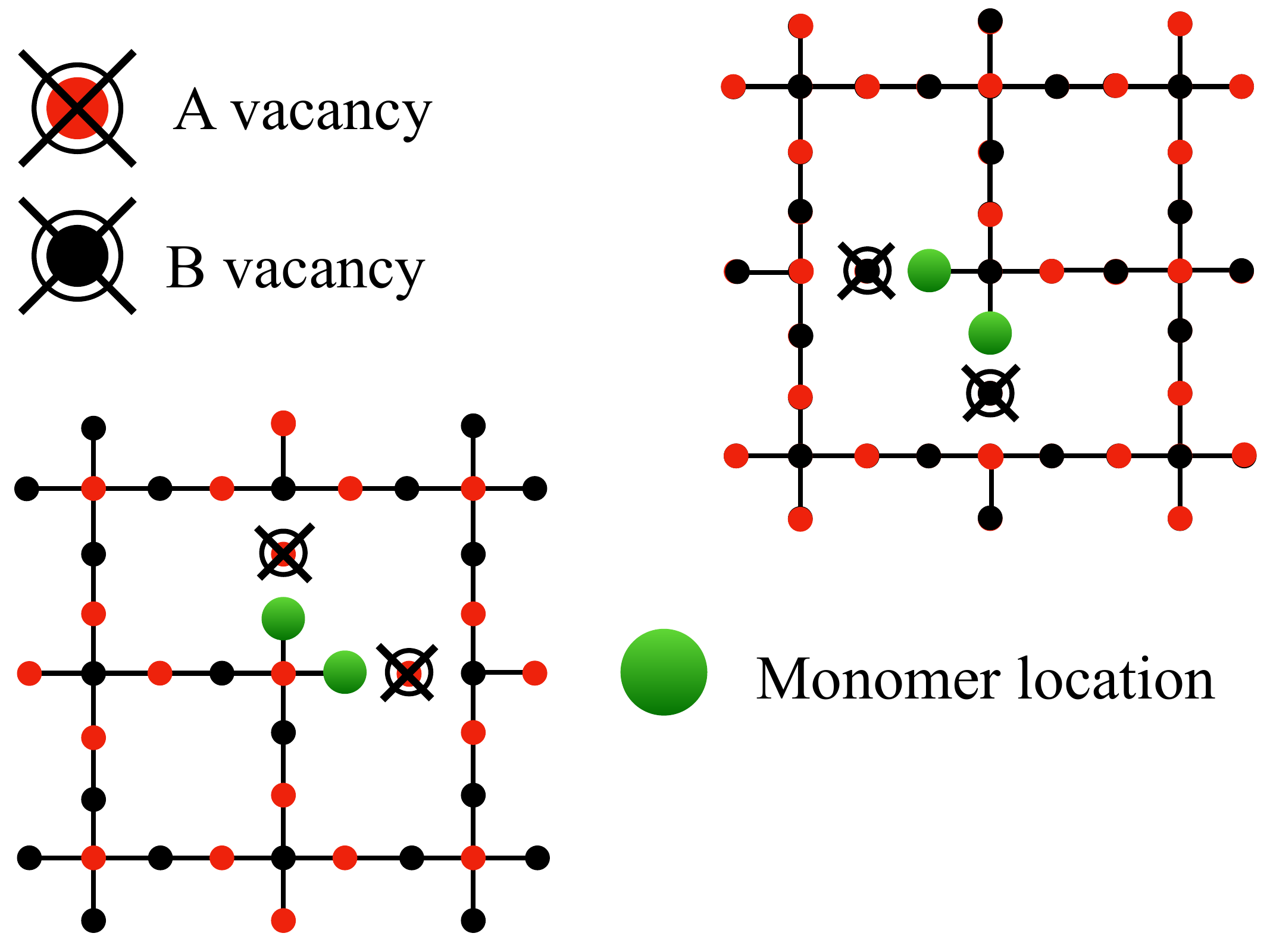}
		\caption[A schematic representation  of two $\mathcal{R}$-type regions, which are created by deleting sites shown as crossed]{A schematic illustration of two $\mathcal{R}$-type regions, which are created by deleting sites shown as crossed. Red denotes the $A$ sublattice, while black denotes the $B$ sublattice. 
			Each $\mathcal{R}$-type region traps a monomer, whose possible locations are shown as the green sites in the illustration. Note that one $\mathcal{R}$-type region is seeded by vacancies on the $A$ sublattice, whereas the other such region is created by deleting some $B$ sublattice sites.			  
		}
		\label{fig:K2latticeRtype}
	\end{figure}

\begin{figure}[h]
		\begin{tabular}{cc}
			\includegraphics[width=0.5\columnwidth]{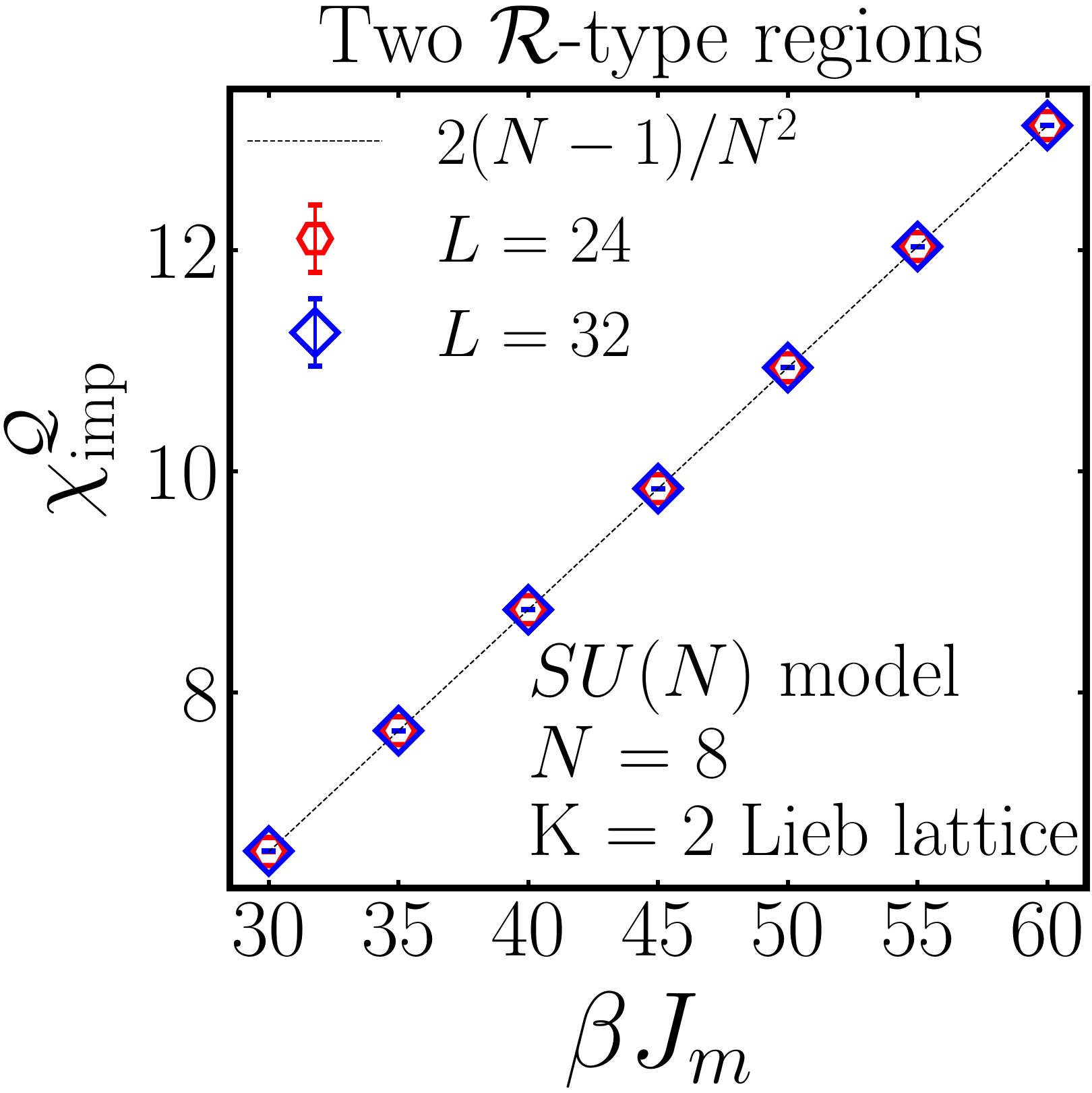} &
			\includegraphics[width= 0.5\columnwidth]{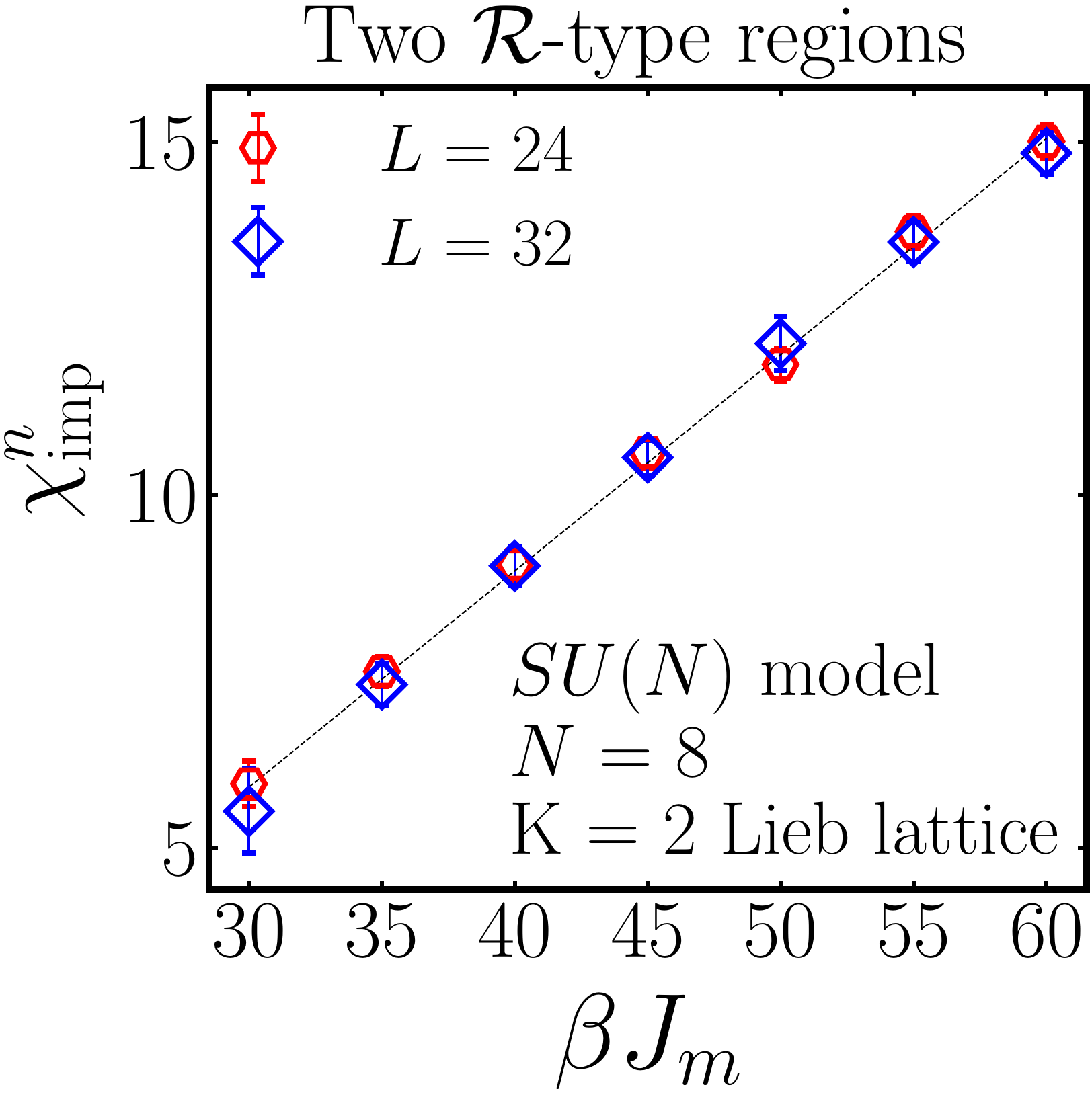}
		\end{tabular}		
		\caption[Impurity susceptibility corresponding to the uniform and alternating sum of the order parameter]{Impurity susceptibilities associated with the response to uniform static fields that couple to  $\mathcal{Q}_{\alpha\alpha}^{\rm tot}$ and $n^{\rm tot}_{\alpha\alpha}$ respectively. Now the data is shown for two $\mathcal{R}$-type regions (depicted in Fig. \ref{fig:K2latticeRtype}) separated by a distance of $L/2$ within an otherwise pure $L \times L$ Lieb lattice with $K = 2$, consisting of $L^2$ unit cells, and periodic boundary conditions applied in both directions.
			Left panel: The impurity susceptibility $\chi^{\mathcal{Q}}_{\rm imp}$ (as defined in the text) for the $K = 2$ Lieb lattice SU($N$) model due to the presence of two such $\mathcal{R}$-type regions shows a clear Curie tail $\propto 1/T \equiv \beta$ at low temperature for values of $N$ that display liquid-like behavior in the $K = 2$ Lieb lattice staggered SU($N$) model.
			Right panel: The impurity susceptibility $\chi_{\rm imp}^{n}$ (as defined in the text) for the $K = 2$ Lieb lattice SU($N$) model due to the presence of two such $\mathcal{R}$-type regions also shows a clearly visible Curie tail $\propto 1/T \equiv \beta$, for values of $N$ that display liquid-like behavior in the $K = 2$ Lieb lattice staggered SU($N$) model.
		}
		\label{fig:K2Rtyperegions}
	\end{figure}

Finally, we examine impurities formed by two monomer-trapping, or $\mathcal{R}$-type, regions on the $K=2$ Lieb lattice. A schematic representation of such a region is shown in Fig.~\ref{fig:K2latticeRtype}, illustrating that each region confines a single monomer on the highlighted sites. We place two such $\mathcal{R}$-type regions at a separation $L/2$ on an $L\times L$ lattice and compute the corresponding impurity susceptibilities. As shown in Fig.~\ref{fig:K2Rtyperegions}, both the uniform (left panel) and staggered (right panel) susceptibilities display clear Curie-tail behavior, signaling the presence of free, localized SU($N$) moments. Moreover, the Curie constant extracted from $\chi^{\mathcal{Q}}_{\rm imp}$ agrees precisely with that of two free SU($8$) ``spins''.

As we demonstrate in Sec.~\ref{subsec:zero_temp}, these emergent moments pair up into long-range singlet valence bonds at zero temperature. This provides further indirect evidence for the existence of free, localized moments bound to the $\mathcal{R}$-type regions, even though isolated vacancies in the same liquid phase do not nucleate such degrees of freedom.

	\subsection{Zero temperature} \label{subsec:zero_temp}

    We begin by examining the honeycomb VBS phase with two isolated vacancies, created by removing one site from the $A$ sublattice and one from the $B$ sublattice. As discussed in Sec.~\ref{subsec:finite_temp}, a vacancy-induced local moment forms in the vicinity of each vacancy in order to avoid the formation of a domain-wall-like defect, which would otherwise incur an energy cost proportional to its length \cite{Levin_Senthil_PRB_2004}. Apart from  the nearest-neighbor exchange coupling 
	$J_{m}$, vacancy-induced local moments also interact with each other via an effective exchange coupling $J_{\rm eff}$, mediated through the inert part of the system. For a gapped quantum paramagnet, $J_{\rm eff}$ is expected to decay exponentially with the distance between the local moments. These local moments are essentially free in the temperature range $J_{\rm eff} \ll T \ll J_{m}$,  resulting in a Curie-tail behavior in the impurity susceptibilities, as confirmed by finite temperature simulations. 
    
    Now, in the low-temperature regime when $T \simeq  J_{\rm eff}$, the impurity susceptibility would start deviating from the Curie-tail behavior, owing to the interaction between the local moments. In an extreme scenario, as the temperature approaches very low values, the impurity susceptibility would eventually saturate due to the formation of a long spin singlet between the local moments, while maintaining the overall honeycomb columnar VBS pattern intact. The formation of a long singlet between the local moments causes the ferromagnetic impurity susceptibility $\chi^{\mathcal{Q}}_{\rm imp}$ to go to zero, whereas the antiferromagnetic one i.e., $\chi^{n}_{\rm imp}$ saturates to a non-zero value. 
	
  The small value of $J_{\rm eff}$ implies that deviations in the impurity susceptibility become observable only at extremely low temperatures (i.e., very large $\beta$). Standard finite-temperature algorithms \cite{Sandvik_JPhysA_1992, Sandvik_PRB_1999, Syljuasen_Sandvik_PRE_2002, Sandvik_AIP_Conference_2010}, however, are not well suited to accessing this low-temperature regime, making a direct numerical observation of such deviations challenging. Fortunately, the projector Monte Carlo method \cite{Sandvik_Evertz_PRB_2010} allows us to overcome this limitation by directly probing the effects of non-magnetic impurities in the paramagnetic ground states of the staggered $SU(N)$ model.

	With this in mind, we compute the expectation of the singlet projector $\hat{P}_{ij}$ defined between two existing ``spin'' degrees of freedom situated on opposite sublattice sites at positions $i$ and $j$. This quantity generally measures the extent to which a spin degree of freedom on the  $A$ sublattice forms a $SU(N)$ singlet with another degree of freedom on the  $B$ sublattice.  The advantage of computing this quantity is two-fold: $(\rm i) $  For the $SU(2)$ model, this expectation value is the two-point spin correlation up to an additive constant. $(\rm ii) $ Even though the projector Monte Carlo simulations crucially rely on a variational wave function resolved into the valence bond basis, the expectation value of this quantity is independent of the basis. 
	\begin{figure}[h]
		\begin{tabular}{c}
			\includegraphics[width=1.0\columnwidth]{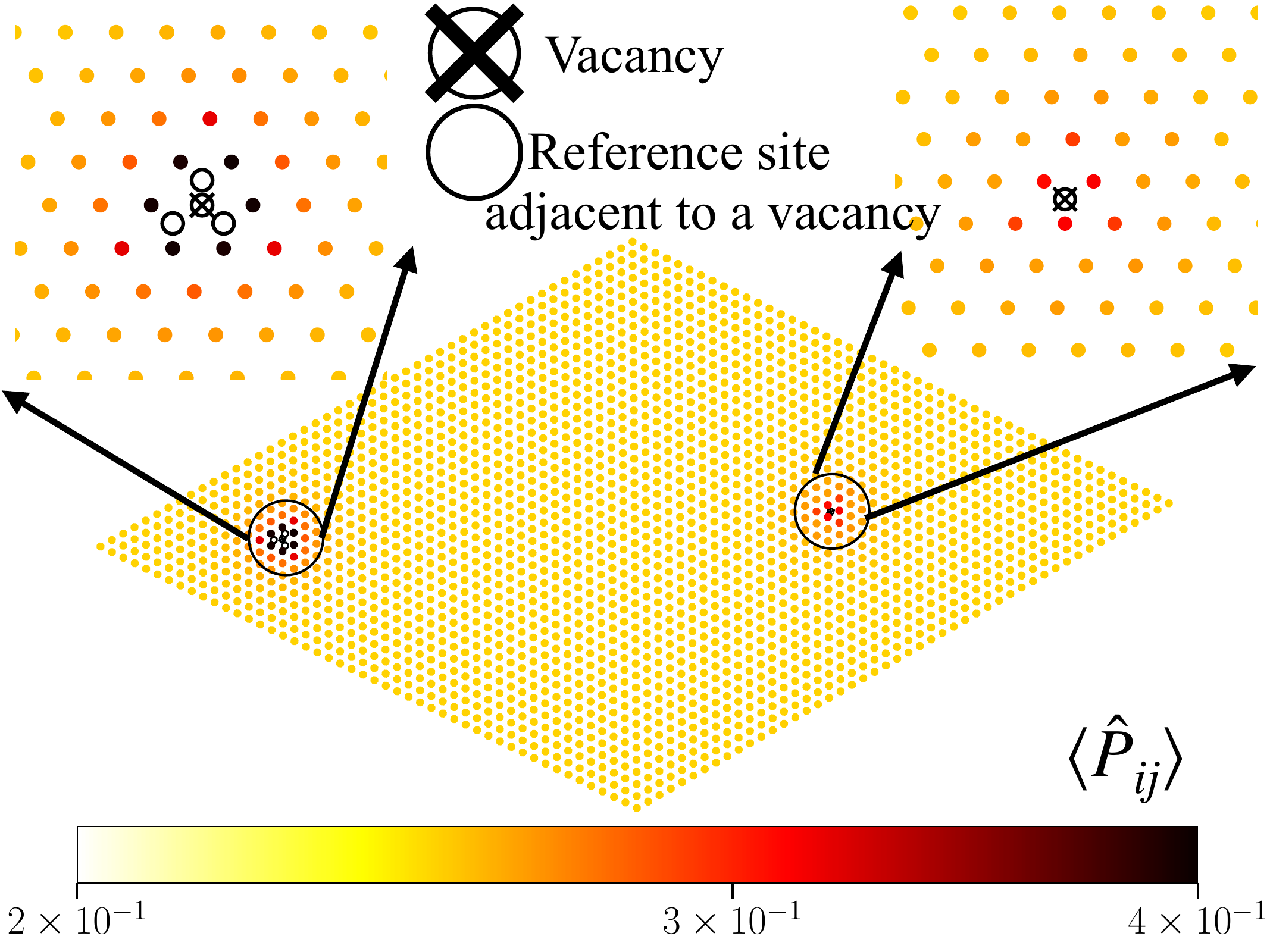} 
		\end{tabular}		
		\caption[Heat map of the singlet projector expectation in the honeycomb VBS phase of the  $SU(2)$ $JQ_3$ model for  $L = 48$]{Heat map of the singlet projector expectation in the honeycomb VBS phase of the  $SU(2)$ $JQ_3$ model for  $L = 48$, with two isolated vacancies at maximal separation. The left side of the heat map zooms in near the first vacancy, while the right side zooms in near the second vacancy. The singlet projector expectation value, $\langle \hat{P}_{ij} \rangle$, is computed for $i$ (shown as the empty circles in the figure) ranging over sites adjacent to the first vacancy and $j$ ranging over all other sites on the opposite sublattice. Despite the large separation between the two vacancies, a distinctly high value of $\langle \hat{P}_{ij} \rangle$ can be observed at the sites $j$ adjacent to the second vacancy. }
		\label{fig:HeatmapHoneycomb}
	\end{figure}
    
	We reiterate that for the honeycomb lattice,  the first vacancy is created by removing a $A$ sublattice site, and the second vacancy is created by removing a $B$ sublattice site. 
	In Fig. \ref{fig:HeatmapHoneycomb}, we display a heat map showing the expectation value of the singlet projector $\langle \hat{P}_{i_{o}j}  \rangle $. 
	This value is calculated from a reference site $i_0$, adjacent to the first vacancy, to all other sites  $j$ on the opposite sublattice.  
	Remarkably, the reference site not only exhibits strong correlations with its neighbors but also demonstrates a prominent correlation with the adjacent sites of the second vacancy, even though the separation between them is large. It's important to note that the sites next to each vacancy in the honeycomb VBS phase are possible locations where local moments might form.
	\begin{figure}[htb!]
		\begin{tabular}{c}
			\includegraphics[width=1.0\columnwidth]{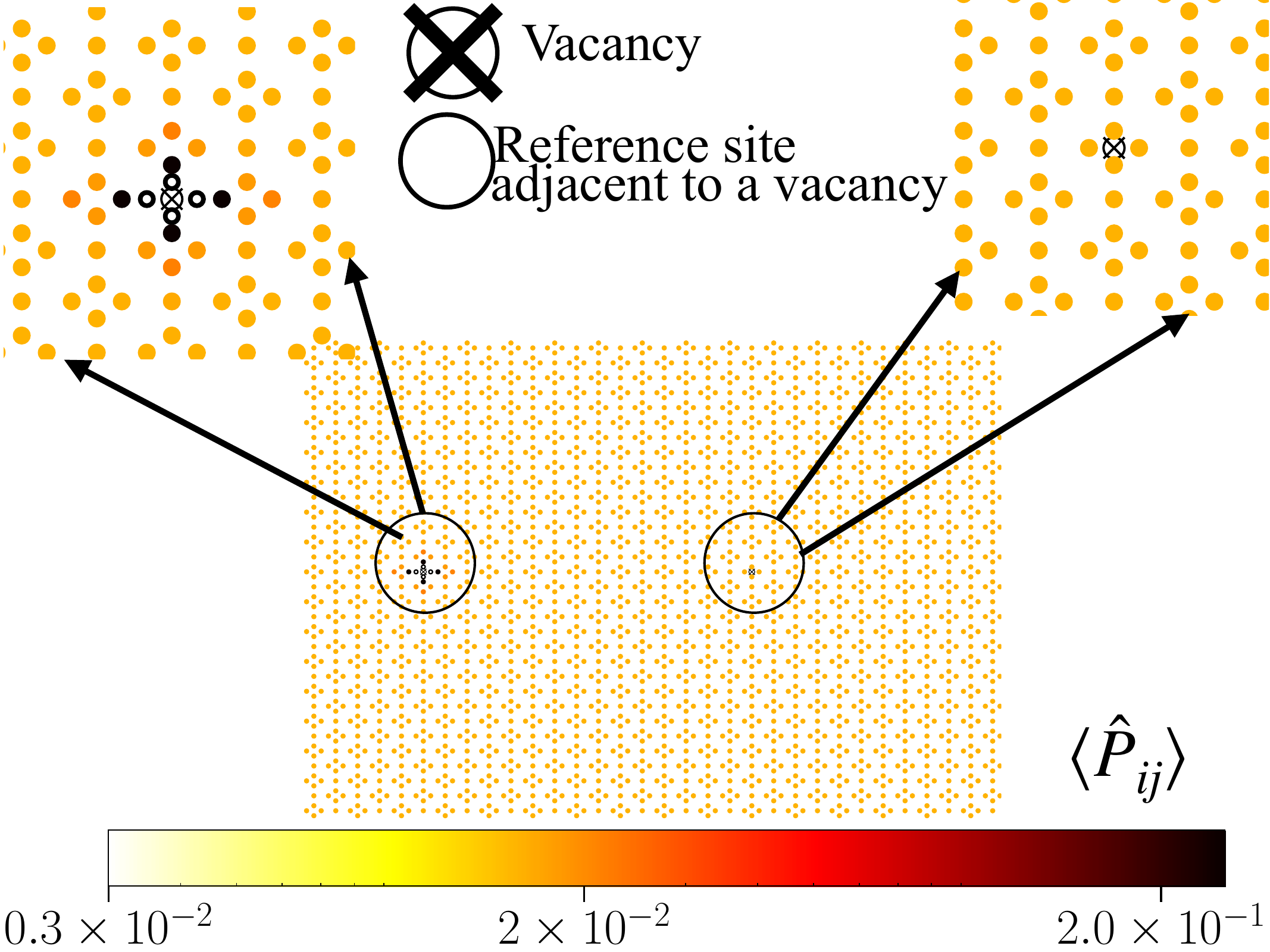} 
		\end{tabular}		
		\caption[Heat map of the singlet projector expectation in the $K = 2$ Lieb lattice liquid regime of the staggered $SU(8)$ model for $L = 32$]{ Heat map of the singlet projector expectation in the $K = 2$ lattice liquid regime of the staggered $SU(8)$ model for $L = 32$, with two isolated vacancies. The left side of the heat map zooms in near the first vacancy, while the right side zooms in near the second vacancy. The singlet projector expectation value, $\langle \hat{P}_{ij} \rangle$, is computed for $i$ ranging over sites(shown as the empty circles in the figure) adjacent to one of the vacancies and $j$ ranging all other sites on the opposite sublattice. The heat map does not exhibit any prominent long-distance values in $\langle \hat{P}_{ij} \rangle$, which aligns with the fact that isolated vacancies do not seed local moments in such liquid states.   }
		\label{fig:HeatmapK2Vacancies}
	\end{figure}
    In contrast, the same computation for the $K = 2$ Lieb lattice, where two isolated single vacancies are created on opposite sublattice sites, does not show a prominent correlation with the sites adjacent to the second vacancy. This observation is consistent with the fact that local moments are absent in a site-diluted gapped RVB liquid phase, which has an exponentially large number of perfect matchings. Fig. \ref{fig:HeatmapK2Vacancies} illustrates this finding.

\begin{figure}[htb!]
		\begin{tabular}{c}
			\includegraphics[width=1.0\columnwidth]{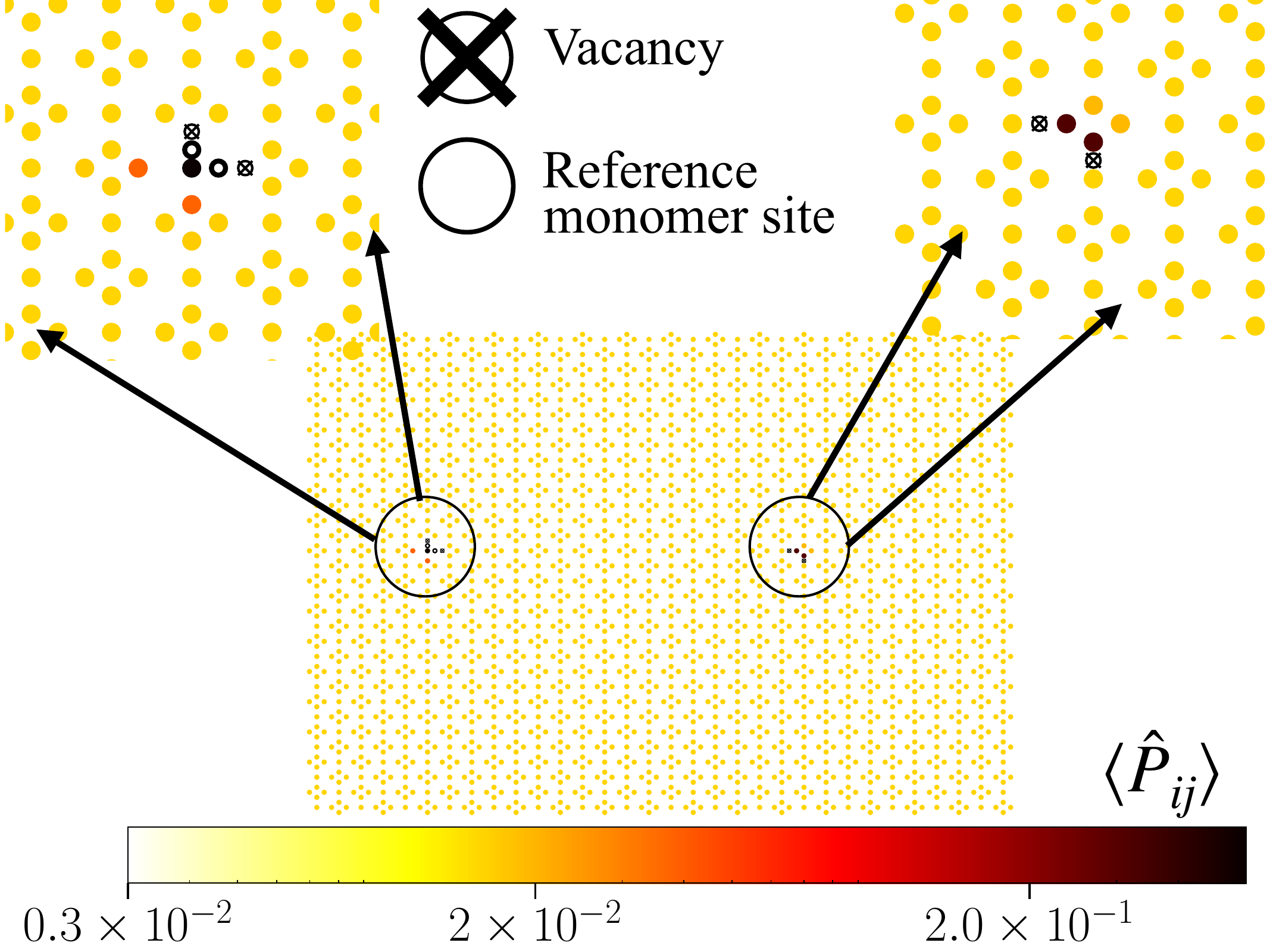} 
		\end{tabular}		
		\caption[Heat map of the singlet projector expectation in the $K = 2 $ Lieb lattice liquid regime of the staggered $SU(8)$ model for $L = 32$]{Heat map of the singlet projector expectation in the $K = 2 $ Lieb lattice liquid regime of the staggered $SU(8)$ model for $L = 32$,  with two isolated $\mathcal{R}$-type regions as shown in Fig. \ref{fig:K2latticeRtype}. The left side of the heat map zooms in near the first $\mathcal{R}$-type region, while the right side zooms in near the second $\mathcal{R}$-type region. 
			The singlet projector expectation value, $\langle \hat{P}_{ij} \rangle$, is computed for $i$ ranging over the monomer sites of the first $\mathcal{R}$-type region and $j$ ranging over all other sites on the opposite sublattice.  Despite the large separation between the two regions, a distinctly high value of $\langle \hat{P}_{ij} \rangle$ can be observed when $j$ is one of the monomer-carrying sites of the second $\mathcal{R}$-type region.
		}
		\label{fig:HeatmapK2Rtype}
	\end{figure}

	Previously, we presented evidence of vacancy-induced local moments upon introducing two maximally separated $\mathcal{R}$-type regions, as depicted in Fig. \ref{fig:K2latticeRtype}, within the liquid-like regime of the $K = 2$ Lieb lattice. Each $\mathcal{R}$-type region hosts a single monomer, representing an unpaired ``spin'' in the valence bond picture and thus indicating a potential source of a vacancy-induced local moment. Indeed, the zero-temperature computation in this system showcases prominent correlations among the monomer sites. This observation is depicted in Fig. \ref{fig:HeatmapK2Rtype}
	\section{Discussion}
	\label{Summary}

	In this work, we have explored the effects of non-magnetic impurities (vacancy disorder) on the quantum paramagnetic phases stabilized by SU($N$) designer Hamiltonians on bipartite lattices. We have provided fairly conclusive evidence that a vacancy-induced local moment forms in the immediate vicinity of each vacancy in a valence bond solid (VBS) phase. We have also identified a spin-liquid-like regime at low temperatures in $SU(N)$ models on the $K = 2$ Lieb lattices. This liquid-like regime serves as a testbed for examining the effects of vacancy disorder in short-range RVB spin liquids. Our results show quite clearlly that isolated vacancies do not seed a local moment in such spin liquids. Instead, vacancy-induced local moments are associated with regions of the disordered lattice that host the monomers of any maximum-density dimer packing of the lattice. Such regions are associated with multi-vacancy clusters, and the formation of such emergent local moments is an essentially multi-vacancy effect in short-range RVB spin liquids of this type. 
	
	It would be interesting to extend these results to obtain spatially-resolved information about the low-energy spectral weight associated with the spinful excitations above the ground state in the presence of vacancies. This could provide an independent window through which one could study the formation of such vacancy-induced local moments. It would also be interesting to study the vacancy-induced susceptibility response at low temperatures in frustrated spin $S=1/2$ models, and use this to diagnose the presence of spin liquid behavior in some part of the phase diagram. We hope our work helps motivate such studies.
	
	\section{Acknowledgements}
	We gratefully acknowledge generous allocation of computing resources by the Department of Theoretical Physics (DTP) of the Tata Institute of Fundamental Research (TIFR), and related technical assistance from K. Ghadiali and A. Salve. The work of MZA and SK was supported at the TIFR by a graduate student fellowship from DAE, India. KD was supported at the TIFR by DAE, India, and in part by a J.C. Bose Fellowship (JCB/2020/000047) of SERB, DST India, and by
	the Infosys-Chandrasekharan Random Geometry Center
	(TIFR). During the drafting of this manuscript, SK was supported at the International Center for Theoretical Sciences of TIFR (ICTS-TIFR) by a postdoctoral fellowship from DAE, India, under project no. RTI4019, while MZA was supported at the National Yang Ming Chiao Tung University, Taiwan, by a postdoctoral fellowship under project no. NSTC 114-2811-M-A49-557.

	\bibliography{Reference}
\end{document}